\documentclass[11pt]{article}
\usepackage{blois,epsfig}
\usepackage[latin1]{inputenc}
\usepackage[OT1]{fontenc}

\def\be{\begin{equation}}
\def\ee{\end{equation}}
\def\bea{\begin{eqnarray}}
\def\eea{\end{eqnarray}}

\begin{document}
\vspace*{4cm}
\title{GZK Horizons and the
Anisotropy of Highest-energy Cosmic Ray Sources}
\author{Chia-Chun Lu$^{1}$ and Guey-Lin Lin$^{1,2}$} \address{$^{1}$Institute of Physics, National
Chiao-Tung University, Hsinchu 300, Taiwan.}\address{$^{2}$Leung
Center for Cosmology and Particle Astrophysics, National Taiwan
University, Taipei 106, Taiwan.}

\maketitle\abstracts{
Motivated by recent Pierre Auger result on the correlation of the
highest-energy cosmic rays with the nearby active galactic nuclei,
we explore possible ultrahigh energy cosmic ray (UHECR) source distributions and their effects on GZK horizons. Effects on GZK horizons by local over-density of UHECR sources are examined carefully with constraints on the degree of local over-density
inferred from the measured UHECR spectrum. We include the energy calibration effect on the Pierre Auger data in our studies. We propose possible local over-densities of UHECR sources which are testable in the future cosmic ray astronomy.}

\section{Introduction}

Recently, Pierre Auger observatory published results on correlation
of the highest-energy cosmic rays with the positions of nearby
active galactic nuclei (AGN) \cite{Abraham:2007bb}. Such a
correlation is confirmed by the data of Yakutsk \cite{Ivanov:2008it}
while it is not found in the analysis by HiRes \cite{Abbasi:2008md}.
In the Auger result, the correlation is maximal for the threshold
energy of cosmic rays at $5.7\times 10^{19}$ eV, the maximal
distance of AGN at $71$ Mpc and the maximal angular separation of
cosmic ray events at $\psi=3.2^{\circ}$. Due to increasing efforts on
verifying the Auger result, it is worthwhile to examine the above
correlation from a phenomenological point of view.

Since the angular scale of the observed correlation is a few
degrees, one expects that these cosmic ray particles are
predominantly light nuclei. The effect of GZK attenuations on these
cosmic ray particles \cite{Greisen,ZK} can be described by a
distance scale referred to as ``GZK horizon". By definition, the GZK
horizon associated with a threshold energy $E_{\rm th}$ is the
radius of a spherical region which is centered at the Earth and
produce $90\%$ of UHECR events arriving on Earth with energies above
$E_{\rm th}$.

Assuming a uniform distribution of UHECR sources with identical
cosmic ray luminosity and spectral index \cite{Harari:2006uy}, the
GZK horizon for protons with $E_{\rm th}=57$ EeV is about $200$ Mpc
while the V-C catalog\cite{VC06} used by Pierre Auger for the
correlation study is complete only up to $100$ Mpc. Such a deviation
may arise from non-uniformities of spatial distribution, intrinsic
luminosity and spectral index of local AGN as mentioned in
\cite{Abraham:2007bb}. In addition, the energy calibration also
plays a crucial role since the GZK horizon is highly sensitive to
the threshold energy $E_{\rm th}$. Energy values corresponding to
the dip and the GZK cutoff of UHECR spectrum were used to calibrate
energy scales of different cosmic ray experiments
\cite{Berezinsky:2008qh,Kampert:2008pr}. It has been shown that all
measured UHECR energy spectra can be brought into good agreements by
suitably adjusting the energy scale of each experiment
\cite{Berezinsky:2008qh}. Furthermore, it has been shown that a
different shower energy reconstruction method infers a $30\%$ higher
UHECR energy than that determined by Auger's fluorescence
detector-based shower reconstruction \cite{Engel:2007cm}.

In this presentation, we report our results \cite{Lu:2008wj} on
examining the consistency between Auger's UHECR correlation study
and its spectrum measurement. The impact by the local over-density
of UHECR sources is studied. We also study the energy calibration
effect on the estimation of GZK horizon and the spectrum of UHECR.
Certainly a $20\%-30\%$ upward shift on UHECR energies reduces the
departure of theoretically calculated GZK horizon to the maximum
valid distance of V-C catalog \cite{Abraham:2007bb}. The further
implications of this shift will be studied in fittings to the
shifted Auger spectrum.

\section{GZK horizons and the UHECR spectrum}
\begin{table}[hbt]
\begin{center}
\caption{GZK horizons of UHECR calculated with the local over-density $n(l<30 {\rm
Mpc})/n_0=1, \ 2, \ 4,$ and $10$, and arrival threshold energy
$E_{\rm th}=57$ EeV, $70$ EeV, $80$ EeV and $90$ EeV respectively. The listed numbers are in
units of Mpc.}
\begin{tabular}{|c|c|c|c|c|}\hline
$n(l<30 {\rm
Mpc})/n_0$ & $E_{\rm th}=57 \, {\rm EeV}$ & $E_{\rm th}=70 \, {\rm
EeV}$ & $E_{\rm th}=80 \, {\rm EeV}$ & $E_{\rm th}=90 \, {\rm EeV}$
\\ \hline $1$ & $220$ & $150$ & $115$ & $90$\\
\hline $2$ & $210$ & $140$ & $105$ & $75$\\
\hline $4$ & $195$ & $120$ & $85$ & $60$\\
\hline $10$ & $155$ & $85$ & $50$ & $30$\\
\hline
\end{tabular}
\end{center}
\label{horizon_p3}
\end{table}
GZK horizons corresponding to different local over-densities and
$E_{\rm th}$ are summarized in Table I.
Within the same $E_{\rm th}$, local over-densities up to $n(l<30
{\rm Mpc})/n_0=4$ do not significantly alter GZK horizons. One could
consider possibilities for higher local over-densities. However,
there are no evidences for such over-densities either from
astronomical observations \cite{overdensity} or from fittings to the
measured UHECR spectrum. We note that GZK horizons are rather
sensitive to $E_{\rm th}$. Table I shows that GZK horizons are $\sim
100$ Mpc or less for $E_{\rm th}\geq 80$ EeV.

Fittings to the Auger spectrum have been performed in
\cite{spec_fit}. In our work, we take into account the over-density
of UHECR sources in the distance scale $l\leq 30$ Mpc. The local
over-density of UHECR sources affects the cosmic-ray spectrum at the
highest energy, especially at energies higher than $5\cdot 10^{19}$
eV. Hence the degree of local over-density can be examined through
fittings to the measured UHECR spectrum.
\begin{table}[hbt]
\begin{center}
\caption{The values of total $\chi^2$ from fittings to the Auger
measured UHECR spectrum. Numbers in the parenthesis are $\chi^2$
values from fittings to the $8$ data points in the energy range
$19.05\leq \log_{10}(E/{\rm eV}) \leq 19.75$. The last $4$ data
points record events with energy greater than $71$ EeV. }
\begin{tabular}{|c|c|c|c|c|}\hline
$n(l<30 {\rm
Mpc})/n_0$ & $1$ & $2$ & $4$ & $10$
\\ \hline $\gamma=2.5$ & $14.12 (9.34)$ & $14.61 (9.93)$ & $17.09 (10.50)$ & $28.09 (13.93)$\\
\hline $\gamma=2.6$ & $16.64 (12.28)$ & $15.56 (11.90)$ & $16.01 (11.83)$ & $20.76 (11.67)$\\
\hline
\end{tabular}
\end{center}
\label{chi_square}
\end{table}

\begin{table}[hbt]
\begin{center}
\caption{The total $\chi^2$ values from fittings to the Auger
measured UHECR spectrum with a $30\%$ upward shift on UHECR
energies. Numbers in the parenthesis are $\chi^2$ values from
fittings to the $8$ data points in the energy range $19.16\leq
\log_{10}(E/{\rm eV}) \leq 19.86$. The last $4$ data points record
events with energy greater than $92$ EeV. }
\begin{tabular}{|c|c|c|c|c|}\hline
$n(l<30 {\rm
Mpc})/n_0$ & $1$ & $2$ & $4$ & $10$
\\ \hline $\gamma=2.4$ & $8.65 (4.30)$ & $7.39 (4.67)$ & $10.26 (6.35)$ & $27.31 (13.34)$\\
\hline $\gamma=2.5$ & $ 11.82 (6.16)$ & $8.67 (5.49)$ & $7.78 (5.23)$ & $16.18 (7.39)$\\
\hline
\end{tabular}
\end{center}
\label{chi_square2}
\end{table}
\begin{figure}[hbt]
\begin{center}
$\begin{array}{cc}
\includegraphics*[width=7.0cm]{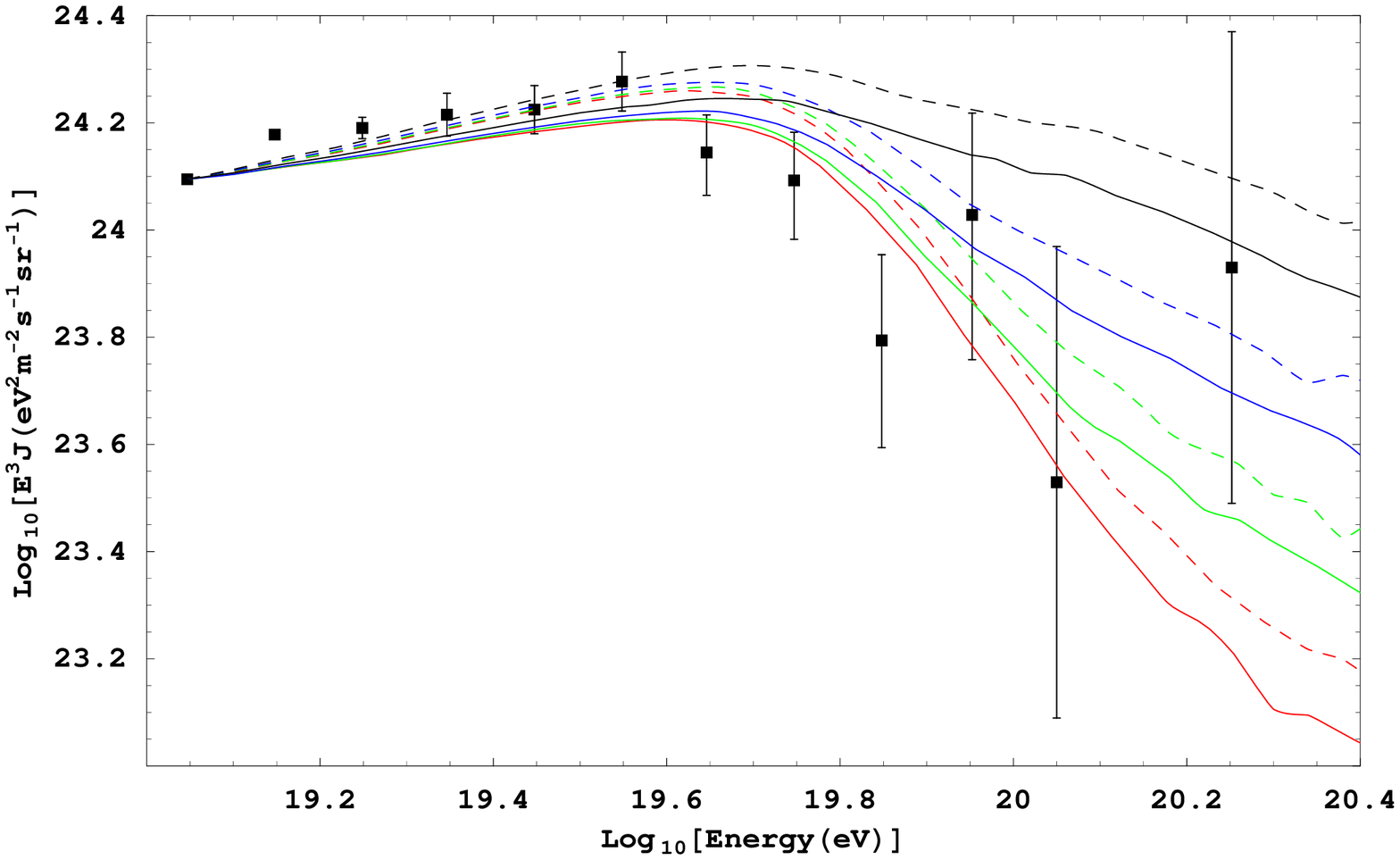} &
\includegraphics*[width=7.0cm]{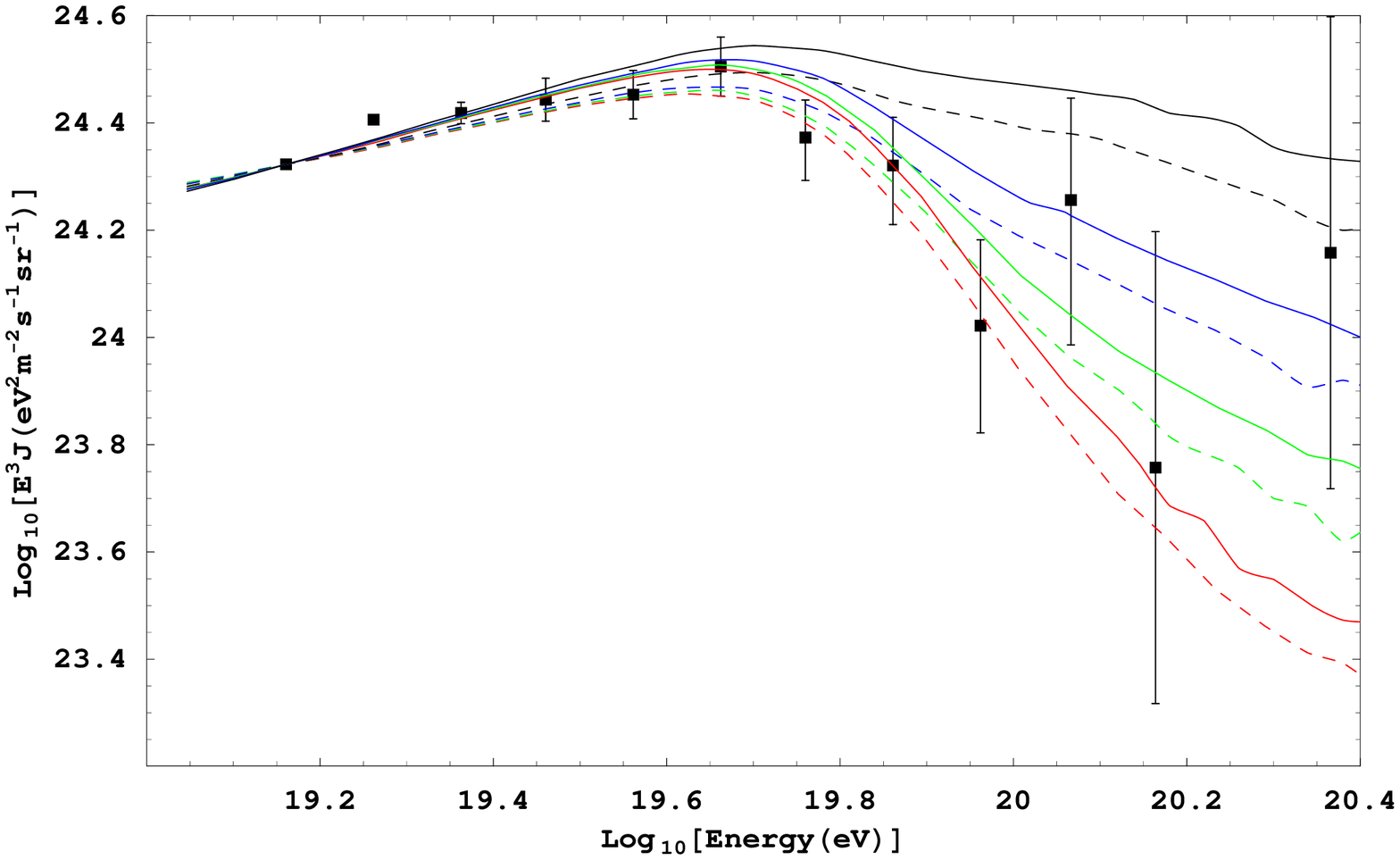}
\end{array}$
\end{center}
\caption{Left and right panels depict fittings to the original and
shifted Auger UHECR spectra respectively with the red, green, blue
and black curves denote models with local over-densities $n(l<30
{\rm Mpc})/n_0=1, \ 2, \ 4,$ and $10$ respectively. Solid and dash
curves correspond to $\gamma=2.6$ and $\gamma=2.5$ respectively. We
take the source evolution parameter $m=3$ throughout the
calculations. } \label{fitting}
\end{figure}
The left paenl in Fig.~\ref{fitting} shows our fittings to the Auger
measured UHECR spectrum with $\gamma=2.5$ and $2.6$ respectively. We
take the red-shift dependence of the source density as
$n(z)=n_0(1+z)^m$ with $m=3$. We have fitted $12$ Auger data points
beginning at the energy $10^{19}$ eV. We make a flux normalization
at $10^{19}$ eV while varying the power index $\gamma$ and the the
degree of local over-density, $n(l<30 {\rm Mpc})/n_0$. Part of
$\chi^2$ values from our fittings are summarized in Table II. We
found that $\gamma=2.5, \ n(l<30 {\rm Mpc})/n_0=1$ gives the
smallest $\chi^2$ value with $\chi^2/{\rm d.o.f.}=1.57$. For the
same $\gamma$, $n(l<30 {\rm Mpc})/n_0=10$ is ruled out at the
significance level $\alpha=0.001$. For $\gamma=2.6$, $n(l<30 {\rm
Mpc})/n_0=10$ is ruled out at the significance level $\alpha=0.02$.
We note that, for both $\gamma=2.5$ and $\gamma=2.6$, the GZK
horizon with $n(l<30 {\rm Mpc})/n_0=10$, $E_{\rm th}=57$ EeV, $m=3$
and $E_{\rm cut}=1000$ EeV is about $155$ Mpc. Since $n(l<30 {\rm
Mpc})/n_0=10$ is clearly disfavored by the spectrum fitting, one
expects a GZK horizon significantly larger than $155$ Mpc for
$E_{\rm th}=57$ EeV.

We next perform fittings to the shifted Auger spectrum. The results
are shown in the right panel in Fig.~\ref{fitting} where the cosmic ray energy is shifted
upward by $30\%$. Part of $\chi^2$ values are summarized in Table
III. The smallest $\chi^2$ value occurs approximately at
$\gamma=2.4$, $n(l<30 {\rm Mpc})/n_0=2$ with $\chi^2/{\rm d.o.f}=
0.82$.
One can see that $\chi^2$ values from current fittings are
considerably smaller than those from fittings to the unshifted
spectrum. Given a significance level $\alpha=0.1$, it is seen that
every local over-density listed in Table III except $n(l<30 {\rm
Mpc})/n_0=10$ is consistent with the measured UHECR spectrum. It is
intriguing to test such local over-densities as will be discussed in
the next section. We note that, with a $30\%$ upward shift of
energies, the cosmic ray events analyzed in Auger's correlation
study would have energies higher than $74$ EeV instead of $57$ EeV.
The GZK horizon corresponding to $E_{\rm th}=74$ EeV is $120$ Mpc
for $n(l<30 {\rm Mpc})/n_0=2$ and $105$ Mpc for $n(l<30 {\rm
Mpc})/n_0=4$.

We have so far confined our discussions at $m=3$. In the literature,
$m$ has been taken as any number between $0$ and $5$. It is
demonstrated that the effect on UHECR spectrum caused by varying $m$
can be compensated by suitably adjusting the power index $\gamma$
\cite{DeMarco:2005ia}. Since GZK horizons are not sensitive to
$\gamma$ and $m$, results from the above analysis also hold for
other $m$'s.

\section{Discussions and conclusions}

We have discussed the effect of local over-density of UHECR sources
on shortening the GZK horizon. The result is summarized in Table I.
It is seen that such an effect is far from sufficient to shorten the
GZK horizon at $E_{\rm th}=57$ EeV to $\sim 100$ Mpc for a local
over-density consistent with the measured UHECR spectrum. With a
$30\%$ energy shift, each cosmic ray event in Auger's correlation
study would have an energy above $74$ EeV instead of $57$ EeV. GZK
horizons corresponding to $E_{\rm th}=74$ EeV then match well with
the maximum valid distance of V-C catalog. Fittings to the shifted
Auger spectrum indicate a possibility for the local over-density of
UHECR sources.

We point out that the local over-density of UHECR sources is
testable in the future cosmic ray astronomy where directions and
distances of UHECR sources can be determined. Table IV shows
percentages of cosmic ray events that come from sources within $30$
Mpc for different values of $E_{\rm th}$ and $n(l<30 {\rm
Mpc})/n_0$. Although these percentages are calculated with
$\gamma=2.4$, $m=3$ and $E_{\rm cut}=1000$ EeV, they are however not
sensitive to these parameters.
\begin{table}[hbt]
\begin{center}
\caption{ Percentages of cosmic ray events originated from sources
within 30 Mpc for different values of $E_{\rm th}$ and $n(l<30 {\rm
Mpc})/n_0$.}
\begin{tabular}{|c|c|c|c|c|}\hline
$n(l<30 {\rm
Mpc})/n_0$ & $E_{\rm th}=57 \, {\rm EeV}$ & $E_{\rm th}=70 \, {\rm
EeV}$ & $E_{\rm th}=80 \, {\rm EeV}$ & $E_{\rm th}=90 \, {\rm EeV}$
\\ \hline $1$ & $0.17$ & $0.27$ & $0.36$ & $0.46$\\
\hline $2$ & $0.30$ & $0.43$ & $0.53$ & $0.63$\\
\hline $4$ & $0.46$ & $0.60$ & $0.70$ & $0.77$\\
\hline $10$ & $0.68$ & $0.79$ & $0.85$ & $0.89$\\
\hline
\end{tabular}
\end{center}
\label{event_dist}
\end{table}
For $E_{\rm th}=57$ EeV and $n(l<30 {\rm Mpc})/n_0=1$, only $17\%$
of cosmic ray events come from sources within $30$ Mpc. For $n(l<30
{\rm Mpc})/n_0=2$ and the same $E_{\rm th}$, $30\%$ of cosmic ray
events are originated from sources in the same region.

In conclusion, we have shown that the deviation of theoretically
calculated GZK horizon to the maximum valid distance of V-C catalog
can not be resolved by merely introducing the local over-density of
UHECR sources. On the other hand, if Auger's energy calibration
indeed underestimates the UHECR energy, such a discrepancy can be
reduced. More importantly, fittings to the shifted Auger spectrum
indicate a possible local over-density of UHECR sources, which is
testable in the future cosmic ray astronomy.

\noindent{\bf Acknowledgements}

We thank A. Huang and K. Reil for helpful discussions. We also thank
F.-Y. Chang, T.-C. Liu and Y.-S. Yeh for assistances in computing.
This work is supported by National Science Council of Taiwan under
the grant number 96-2112-M-009-023-MY3.

\end{document}